\documentclass[conference]{IEEEtran}
 \usepackage{fancyhdr} 
\usepackage[T1]{fontenc}
\usepackage{cite}
\usepackage{amsmath,amssymb,amsfonts}
\usepackage{algorithmic}
\usepackage{graphicx}
\graphicspath{{figures/}}
\usepackage{textcomp}
\usepackage{xcolor}
\usepackage{booktabs}
\usepackage{multirow}
\usepackage{hyperref}
\usepackage{subcaption}
\usepackage{listings}
\usepackage{balance}
\usepackage{tikz}
\usepackage{svg}
\usepackage{comment}
\usetikzlibrary{positioning, arrows.meta, calc}

\def\BibTeX{{\rm B\kern-.05em{\sc i\kern-.025em b}\kern-.08em
    T\kern-.1667em\lower.7ex\hbox{E}\kern-.125emX}}
\begin{document}

\title{Complexity and Scale in AI-Assisted Workflow Management: A Federated Learning Case Study}

\author{
  \IEEEauthorblockN{Komal Thareja\IEEEauthorrefmark{1},
    Hamza Safri\IEEEauthorrefmark{2},
    Rajiv Mayani\IEEEauthorrefmark{2},
    Anirban Mandal\IEEEauthorrefmark{1},
    Ewa Deelman\IEEEauthorrefmark{2}}
  \IEEEauthorblockA{\IEEEauthorrefmark{1}RENCI, University of North Carolina at Chapel Hill, NC, USA}
  \IEEEauthorblockA{\IEEEauthorrefmark{2}Information Sciences Institute, University of Southern California, Marina del Rey, CA, USA}
}

\maketitle
\thispagestyle{fancy}
\lhead{}
\rhead{}
\chead{}
\lfoot{\footnotesize{
SC26 Workshops, November 15-20, 2026, Chicago, Illinois, USA
\newline 979-8-3195-1221-5/26/\$31.00 \copyright 2026 IEEE}}
\rfoot{}
\cfoot{}
\renewcommand{\headrulewidth}{0pt}
\renewcommand{\footrulewidth}{0pt}
\begin{abstract}
Federated learning over medical images is a demanding workflow application. Each round fans out across parallel client jobs and converges on an aggregation step that feeds the next round. At scale this yields 101 sub-workflows and 2,679 jobs on GPUs at four sites, which takes an expert months to build, mostly on workflow mechanics rather than science. We ask how far AI assistance can automate such workflows. An LLM agent, grounded in a released plugin of Pegasus-specific skills, first produces a reviewable specification of checkable constraints and acceptance criteria, then generates the executable workflow. A validation loop repairs runtime failures, checks code against those constraints, and regenerates the implementation from the specification alone. We evaluate three LLM agents, report end-to-end runs on the FABRIC testbed, and show how conformance checking against the specification caught three silent errors that failure-driven debugging missed, including one that trained 1,700 jobs on random tensors.
\end{abstract}


\section{Introduction}

Scientific workflows provide a structured means of coordinating computational tasks, data dependencies, and execution resources in complex scientific applications. As applications become increasingly distributed and data intensive, translating scientific requirements into executable workflows requires expertise beyond the application itself, including task decomposition, data movement, dependency management, and execution across heterogeneous resources.

These challenges arise naturally in federated learning (FL), where distributed computation must be coordinated across multiple data holders and successive training rounds. FL trains a shared model without centralizing data~\cite{mcmahan2017communication}, making it particularly relevant to domains such as medical imaging. Its computation is both parallel and iterative: clients train locally within each round, their updates are aggregated, and the resulting global model becomes the input to the next round. Repeating this process across clients, datasets, and execution sites creates a hierarchical workflow with dependencies that extend across rounds.

Workflow management systems (WMSs) such as Pegasus~\cite{deelman2015pegasus}, Nextflow~\cite{di2017nextflow}, and Galaxy~\cite{afgan2018galaxy} provide planning, data staging, orchestration, provenance, and fault recovery, but constructing such workflows still requires substantial system-specific expertise. For this application, the complexity lies less in the learning algorithm than in workflow mechanics, including sub-workflow composition, cross-round data dependencies, catalog propagation, cleanup policies, container management, and data staging. A comparable Pegasus FL workflow developed manually required an estimated 3 to 4 months, with most of the effort devoted to workflow mechanics~\cite{safri2024workflow}.

Automating workflow authoring is therefore attractive. Prior approaches include semantic planning systems such as Wings~\cite{gil2007wings}, which provide correctness guarantees but require manually engineered representations, and more recent LLM-based approaches that generate workflows from natural-language descriptions~\cite{zeng2024flowmind,zhang2024aflow}. These approaches show that user intent can be translated into executable pipelines, but direct code generation leaves the intended design implicit in the implementation, so it is hard to review the design before execution, check the code against its requirements, or regenerate the implementation independently of the original interaction.

In this work, {\em we investigate a specification-driven approach to AI-assisted scientific workflow management, and demonstrate how AI assistance can automate the construction, execution and debugging of complex hierarchical workflows} (e.g. federated learning). The LLM first translates user intent into a structured, reviewable specification containing workflow constraints and acceptance criteria and then generates the executable workflow. Workflow-specific knowledge is provided through \texttt{pegasus-ai}, a reusable plugin of Pegasus-specific skills~\cite{repo_claude_plugin}. The specification remains an independent artifact that can subsequently be used for conformance checking and regeneration, while a validation loop combines these checks with failure-driven debugging during execution.

The paper makes three contributions:
(1) an end-to-end study of {\em AI-assisted composition and management of a complex scientific workflow, from natural-language intent through specification, code generation, debugging, validation, and execution at scale},
(2) an {\em evaluation of multiple LLM coding agents for workflow construction and debugging}, compared with a previously reported expert implementation and
(3) an {\em analysis of defects encountered during development that distinguishes failures accessible to runtime debugging from silent correctness violations}, motivating the combination of failure-driven debugging and {\em specification-based conformance checking for complex workflows}.

% ============================================================
\section{Federated Learning Workflow Complexity}
 \label{sec:application}

\begin{figure}[t]
    \centering
    \includegraphics[width=0.78\linewidth]{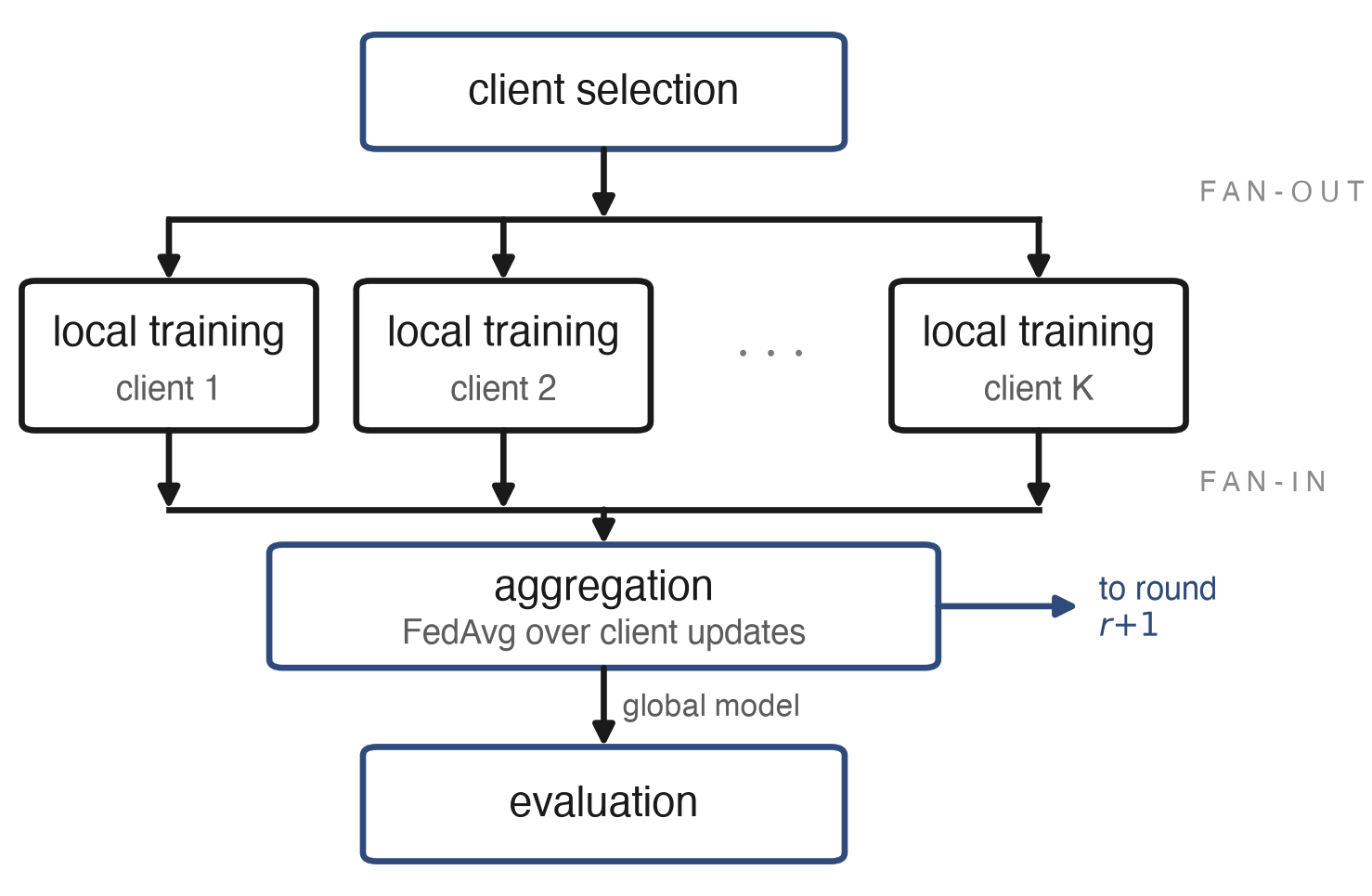}
    %\caption{One federated learning round. Fan-out to parallel client training, fan-in to an averaged global model, which starts the next round.}
    \caption{One federated learning round: parallel client training, followed by aggregation into the global model for the next round.}
    \label{fig:flround}
\end{figure}

Federated learning combines distributed model training with an iterative computational structure that becomes increasingly complex as the number of clients and training rounds grows. Within each round (Fig.~\ref{fig:flround}), a selection step determines the participating clients, which train locally in parallel on their private data shards. An aggregation job then combines the resulting model updates using Federated Averaging (FedAvg)~\cite{mcmahan2017communication}, and an evaluation job assesses the new global model. A complete study repeats this process for $T$ rounds over each dataset, with data preparation at the beginning and centralized baseline training and figure generation at the end.

Three properties make this application challenging to express as a scientific workflow. First, the graph is \emph{iterative}: its size grows with the number of rounds, while the same round structure is repeatedly instantiated. Second, it is \emph{hierarchical}, because a round is itself a natural unit of composition rather than an individual job. Third, it is \emph{dependency-intensive across these levels}, since the global model produced within one round is consumed by the next. In Pegasus, this structure maps naturally to the SubWorkflow construct (sub-workflow hereafter), where each round is represented as a self-contained workflow composed into a parent graph and planned when the parent reaches it. This organization provides per-round fault tolerance, provenance, and cleanup, but relies on advanced workflow-management mechanisms that non-expert users rarely employ.

The difficulty of constructing this workflow lies largely in the mechanics surrounding this structure rather than in the learning algorithm itself. A sub-workflow planned at runtime requires its replica, transformation, and site catalogs to be propagated correctly. Cleanup policies must preserve the global model checkpoint required by the subsequent round, and client shards must remain registered so that they are available when the next round is planned. Large training containers must be reused rather than repeatedly staged, while concurrent experimental configurations must be namespaced to prevent results from being overwritten. Each of these issues occurred during the development of the workflow and is examined further in Section~\ref{sec:debugging}.

% ============================================================
%\section{AI-Assisted Workflow Generation}
\section{AI-Assisted Workflow Management}

\subsection{Overview of AI-Assisted Workflow Generation}
\begin{figure}[h]
    \centering
    \includegraphics[width=\columnwidth]{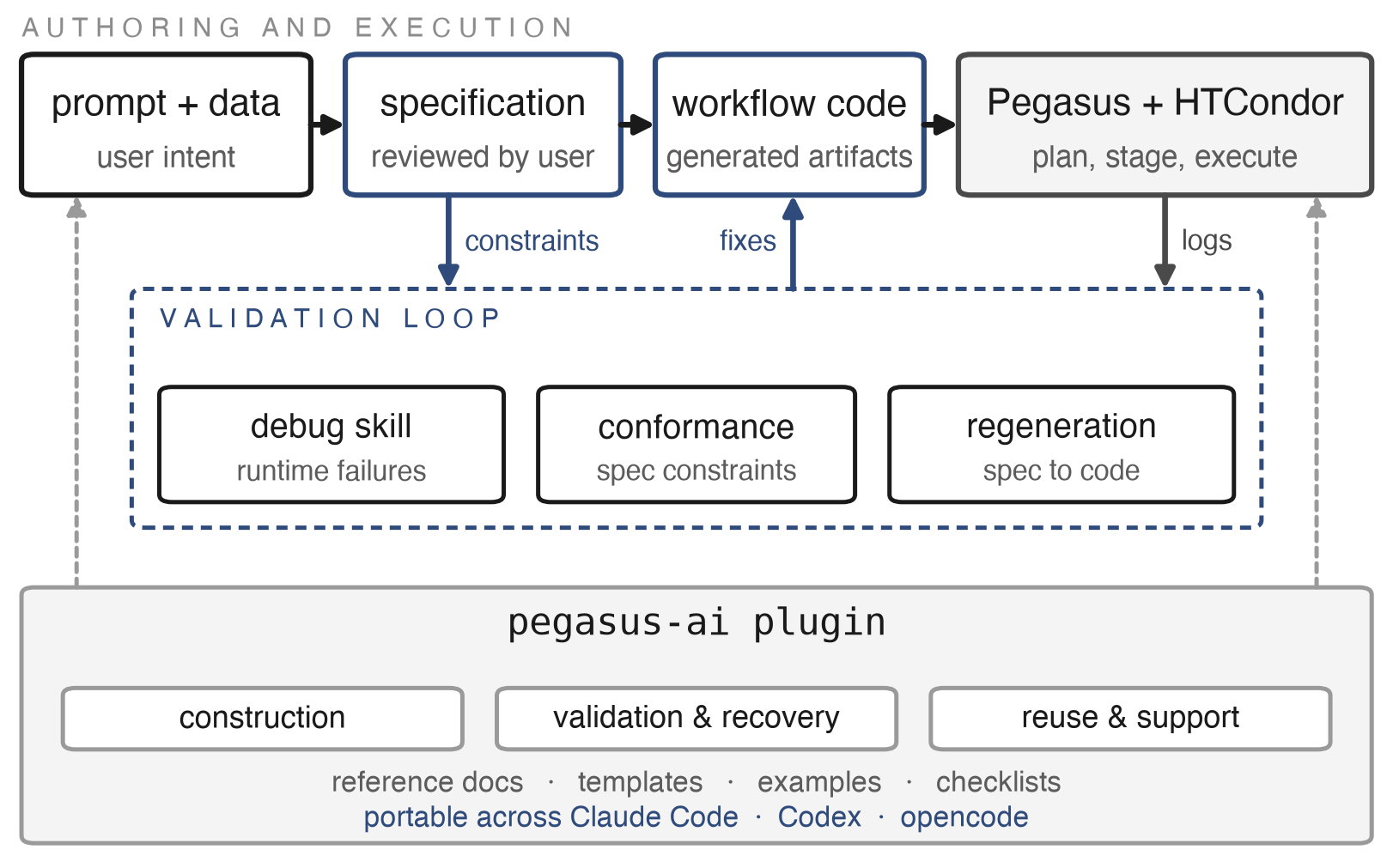}
    \caption{AI-assisted workflow management.}
    \label{fig:flow}
\end{figure}

We propose a specification-driven approach to AI-assisted workflow generation, decomposed into three explicit stages that separate user intent from implementation (Fig.~\ref{fig:flow}). A \textbf{prompt and dataset description} captures high-level intent, constraints, and experimental objectives without requiring knowledge of the underlying workflow system. From it the system generates a \textbf{structured specification} that formalizes the design as a human-readable artifact the user validates and refines before anything is executed. Once the specification is approved, \textbf{code generation} produces directly deployable artifacts: workflow definitions, wrapper scripts, configuration files, and container specifications. \textbf{Workflow-specific skills} ground the model in reusable workflow knowledge and templates throughout, so that outputs conform to the requirements of the target execution environment.

Each participant has an explicit role. The \emph{user} provides the prompt and dataset description, reviews and approves the specification, and decides among design alternatives that the system surfaces, such as the orchestration strategies of Section~\ref{sec:ai_assisted}. The \emph{LLM} interprets the prompt, synthesizes the specification, and generates the artifacts. The \emph{skills} are authored by workflow-system developers and supply the Pegasus-specific knowledge that grounds the model. \emph{Pegasus} plans and orchestrates the workflow and \emph{HTCondor} executes its jobs, and the LLM modifies neither.

The methodology is \textbf{tool-agnostic}, assuming only an LLM agent capable of invoking structured skills. Claude Code with the \texttt{pegasus-ai} plugin is the reference implementation here, and Section~\ref{sec:results} evaluates it against agents that receive the same workflow knowledge as unstructured context.

% ============================================================

\subsection{Workflow-Specific Plugins and Skills}

We organize workflow knowledge into \emph{plugins}, packaged extensions that supply the language model with workflow-specific material, and within each plugin into \emph{skills}, each implementing one workflow-oriented operation. A skill is a self-contained, agent-agnostic slash command bundling instructions with the reference material that grounds them: documentation, example workflows, validation checklists, and reusable templates. The model invokes only the skill the current step needs, which holds generation to Pegasus-specific requirements such as correct API usage, data staging patterns, and container configuration.

The plugin we built for this work is \texttt{pegasus-ai}~\cite{repo_claude_plugin}, published in a marketplace of such plugins for scientific workflow development. Its eight skills group into three functional categories: workflow construction, validation \& recovery, and reuse \& support. Any agentic coding tool supporting the skill and plugin interface, such as Claude Code, Codex, or opencode, can load it and invoke the same skills.

\paragraph{Workflow construction}
The \texttt{/pegasus-scaffold} skill generates a complete workflow project from a high-level pipeline description, including the workflow generator, wrapper scripts, Dockerfile, and site catalogs. The \texttt{/pegasus-wrapper} skill creates wrapper scripts that interface Pegasus job execution with computational tools. The \texttt{/pegasus-dockerfile} skill generates container definitions with the required software dependencies for reproducible execution.

\paragraph{Validation and recovery}
The \texttt{/pegasus-debug} skill drives the diagnose, fix, and re-run loop described in Section~\ref{sec:architecture}, while the \texttt{/pegasus-review} skill inspects workflow code for common anti-patterns and correctness issues.

\paragraph{Reuse and support}
The \texttt{/pegasus-convert} skill converts workflows between supported formats. The \texttt{/pegasus-help} skill provides context-aware Pegasus documentation and guidance. The \texttt{/kiso} skill generates experiment configurations for the Kiso \cite{mayani2026kiso} experiment management platform.

\paragraph{How the plugin was developed} The \texttt{pegasus-ai} plugin was distilled from building fourteen real Pegasus workflows~\cite{thareja2026sensors, repo_pegasus_workflows} across varied workflows and input modalities, taking roughly $190$ prompts over about $15$ developer-days. Whenever a prompt recurred or a Pegasus-specific pitfall reappeared, it became a reusable skill carrying the documentation, examples, and validation checklists needed to avoid it next time. That authoring cost is amortized over every later workflow and is separate from the per-workflow effort reported in Section~\ref{sec:results}. Because the skills capture Pegasus mechanics rather than domain logic, new scientific domains are usually handled by existing skills, and only a new capability such as targeting a different workflow system would require a new one.

\subsection{Specification-Driven Development}
\label{sec:spec_driven}

Our approach uses a specification-driven development methodology \cite{piskala2026spec} for workflow generation. Rather than directly synthesizing executable code from user input, the system first produces a structured, human-readable specification that captures the workflow design (data dependencies, execution stages, and configuration parameters), together with a validation plan (experiments, metrics, and expected figures), the design decisions resolved with their trade-offs, and an implementation roadmap. This intermediate artifact is reviewed before implementation, and its experiment plan and smoke-test configuration serve as acceptance checks for the generated code.

To make the specification checkable rather than purely descriptive, it distinguishes three kinds of validation content. \emph{Constraints} state properties the implementation must satisfy, such as the round structure of the DAG, the staging of the global model between rounds, and the resource limits of the target execution sites. \emph{Non-constraints} record degrees of freedom deliberately left to the implementation, such as the naming of intermediate files or the internal structure of wrapper scripts, so that the review focuses on decisions that matter and later changes to these aspects do not require re-approval. \emph{Expected outcomes} define acceptance criteria for the generated code, including the artifacts each stage must produce, the smoke-test configuration exercised before full-scale runs, and the metrics and figures anticipated by the experiment plan. User review therefore evaluates concrete, testable statements rather than prose intent, and the same statements later serve as the reference for automated debugging (Section~\ref{sec:debugging}).

For the federated learning case study in this paper, the generated specification is a fourteen-section, $\sim$7,000-word document (Table~\ref{tab:spec}). Validation is performed through user review before any code is generated: the user checks the stated constraints and expected outcomes, confirms the DAG structure, selects among the surfaced orchestration options, and approves the experiment plan. The specification remains a living artifact: its final two sections record implementation status and the fixes applied during debugging (Section~\ref{sec:debugging}), so the persisted document reflects the workflow as executed.

\begin{table}[h]
\centering
\caption{Section outline of the generated specification.}
\label{tab:spec}
\scriptsize
\begin{tabular*}{\columnwidth}{@{\extracolsep{\fill}}ll@{}}
\toprule
1. Project Overview & 8. Experiment Plan \\
2. Datasets & 9. Directory Structure \\
3. Federated Learning Design & 10. References \\
4. Pegasus Workflow Architecture & 11. Design Decisions (Resolved) \\
5. Software Stack \& Containers & 12. Setup \& Dependencies \\
6. Catalogs & 13. Implementation Status \\
7. Execution Environment & 14. Key Fixes Applied \\
\bottomrule
\end{tabular*}
\end{table}

The specification is intentionally a design-level artifact rather than a formal workflow language. %Formal representations such as the Common Workflow Language (CWL)~\cite{crusoe2022methods} define executable semantics for a workflow that already exists, whereas the specification captures the design space \emph{before} implementation: intent, alternatives considered, resolved trade-offs, and acceptance criteria, none of which are expressible in an executable workflow description. The two are complementary: code generation produces a fully formal Pegasus workflow definition, and reproducibility derives from regenerating that definition from the persisted specification.
Only after validation does the system generate the executable artifacts. Where direct LLM interaction leaves the design implicit in whatever code the last prompt happened to produce, this separation makes the design explicit, reviewable, and auditable.

\subsection{Execution, Debugging, and Validation}
\label{sec:architecture}

Generated artifacts are submitted to Pegasus, which constructs the workflow DAG, manages data staging, coordinates job execution, and monitors progress. Execution is delegated to HTCondor~\cite{thain2005distributed}, which handles resource allocation, job scheduling, and distributed execution with built-in fault tolerance. Both run on the submit node alongside the coding agent, and neither is modified by it (Fig.~\ref{fig:flow}).

The system includes an \textbf{AI debugging capability}, delivered by the \texttt{/pegasus-debug} skill and executed by the host coding agent, that handles failure diagnosis and recovery during workflow execution. Pegasus and HTCondor generate logs and runtime metadata as workflows execute. The skill interprets those traces and matches them to known failure patterns such as file staging errors, container issues, resource misconfiguration, dependency conflicts, and incorrect wrapper scripts. It then acts on its own diagnosis: the host agent edits the affected workflow definitions, wrapper scripts, and configuration files and re-submits the affected jobs or sub-workflows, forming a \textbf{closed-loop recovery mechanism} in which failures are detected, diagnosed, and corrected during execution.

Debugging alone is reactive, so it sits inside a wider \textbf{validation loop} that plays the role continuous integration plays in software development, adapted to an artifact that cannot simply be unit tested because one run of it occupies nine GPUs for a day. The loop has three parts, applied at different points in the lifecycle. The debugging skill acts during execution and repairs what fails. A \emph{conformance pass} acts on the code before submission and checks it against the constraints the specification states, which requires no execution at all. A \emph{regeneration test} acts on the specification and rebuilds the implementation from it alone, which is what keeps the specification honest as the code changes underneath it. Section~\ref{sec:results} reports all three. Only the first is driven by failure, and Section~\ref{sec:debugging} shows that this distinction, rather than the sophistication of any one part, determines which defects are reachable at all.

% ============================================================
\section{Evaluation Setup}
\label{sec:experiments}

\subsection{Datasets}

\textbf{TCIA}~\cite{clark2013cancer} provides 4,144 3D CT/MRI lung volumes across multiple collections. One representative slice per imaging series is extracted, resized to $224 \times 224$, and distributed across $K$ clients. The task is binary classification for lung pathology present or absent. Labels are assigned at the collection level from collection metadata, and the train/test split is performed at the series level after pooling all collections.

\textbf{NIH ChestX-ray}~\cite{wang2017chestx} consists of 50,000 2D chest X-rays resized to $224 \times 224$, partitioned by patient ID across $K$ clients, with binary classification (``No Finding'' vs.\ pathology). The held-out test set is reserved at the patient level, so no patient contributes to both training and evaluation.

\subsection{Experimental Setup}

The experimental workflows are generated using the approach described in Section~\ref{sec:architecture}. All experiments use a ResNet-18 model with an ImageNet-pretrained backbone, batch size 32, learning rate 0.001, and stochastic gradient descent with momentum 0.9 and weight decay $10^{-4}$.

We report three configurations that explore scalability and communication efficiency (Table~\ref{tab:experiments}). All $K$ clients participate in every round (client fraction $C=1$). The per-round selection step is seeded by the round number, so a re-planned or rescued round selects the same clients.

E1 is the FedAvg reference against which the others vary a single factor. E3 halves the client count, which increases the data each client holds while reducing aggregation diversity. E4 trades rounds for local epochs. Communication cost here is the number of full-model exchanges between clients and aggregator. With full participation each round moves the global model to and from every client (about 45\,MB per direction), so bytes transferred scale with $K \times T$ and E4 moves one fifth of E1's bytes. Communication efficiency is the accuracy reached for that cost. Numbering follows the released artifact~\cite{repo_claude_fl}, which defines further configurations outside the scope of this evaluation. 

Experiments ran on a distributed HTCondor pool on the FABRIC testbed~\cite{baldin2020fabric} with four GPU-equipped worker nodes across sites, HAWI (3~GPUs), MAX-1 (2), MAX-2 (2), and NCSA (2), totaling 9 GPUs. Jobs ran in Docker containers using CondorIO for data staging, each training job requesting 1 GPU, 4 CPU cores, and 15\,GB memory.

\begin{table}[t]
\centering
\caption{Experimental configurations.}
\label{tab:experiments}
\scriptsize
\begin{tabular*}{\columnwidth}{@{\extracolsep{\fill}}llcccl@{}}
\toprule
\textbf{Exp} & \textbf{Alg.} & \textbf{R} & \textbf{K} & \textbf{E} & \textbf{Focus} \\
\midrule
E1 & FedAvg & 50 & 10 & 2 & Baseline \\
E3 & FedAvg & 50 & 5  & 2 & Scalability \\
E4 & FedAvg & 10 & 10 & 5 & Communication \\
\bottomrule
\end{tabular*}

\end{table}

\subsection{Evaluation Metrics}

\textbf{Workflow generation} is assessed along four dimensions. Execution success measures completion without runtime errors. Structural quality measures whether the generated code satisfies the constraints stated in the specification. Reproducibility measures whether the workflow can be regenerated from the persisted specification alone. Development effort measures the iterations and manual interventions needed to obtain a working workflow. Structural quality is measured by a conformance pass in which ten independent reviewer sessions each receive one copy of the workflow together with the specification, with no access to the development history and no pointer to any particular constraint, and return a verdict of pass, fail or unclear for each of the eleven stated constraints along with the file and line supporting it. A verdict of fail counts as a false positive unless the violation can be independently confirmed in the code.

\textbf{Federated learning performance} is assessed with accuracy and F1 score, only to confirm that the workflows produce interpretable machine learning results.

% ============================================================
\section{Results}
\label{sec:results}

\subsection{Workflow Design and Generation}
\label{sec:ai_assisted}

During the specification phase, the LLM presented two orchestration strategies for the iterative FL rounds: (1)~an \emph{ensemble-manager} pattern, in which a single top-level workflow re-invokes itself across rounds using an external coordinator, and (2)~a \emph{sub-workflow} pattern, in which each round is encapsulated as a Pegasus sub-workflow with explicit data dependencies between rounds. The ensemble-manager approach offers lower scheduling overhead and supports dynamic runtime decisions such as early stopping, but introduces a single point of failure and coarsens fault tolerance and provenance. The sub-workflow approach costs additional per-round scheduling latency and buys back the per-round guarantees of Section~\ref{sec:application}. The user selected the sub-workflow approach, as these benefits outweigh the overhead for long-running FL workflows where GPU computation dominates execution time.

\begin{figure}[h]
    \centering
    \includegraphics[width=\columnwidth]{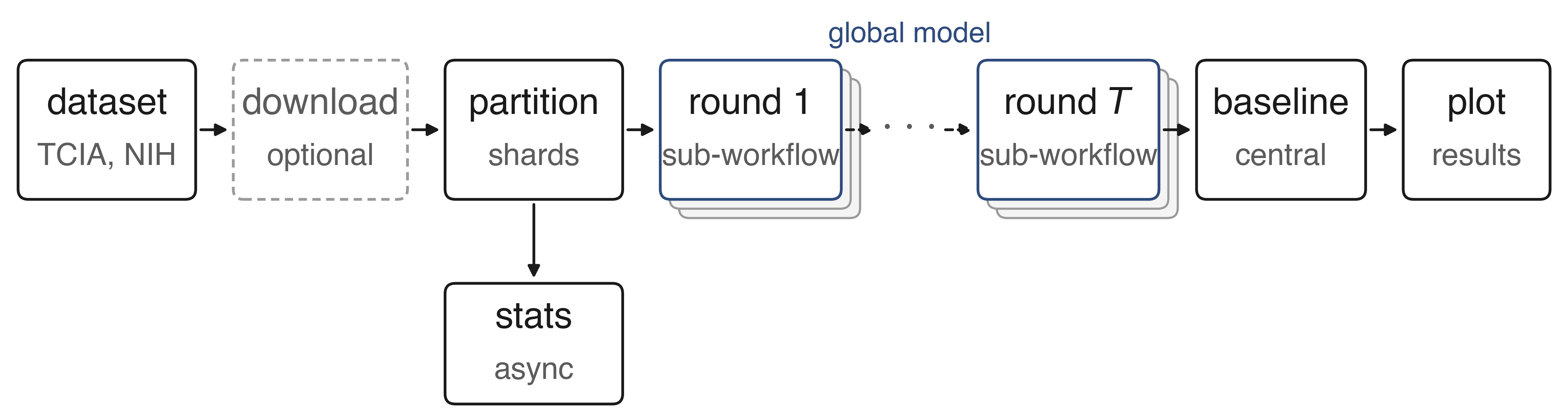}
    \caption{Generated top-level DAG for one dataset branch. Each round is a sub-workflow of the form in Fig.~\ref{fig:flround}.}
    \label{fig:workflow_dag}
\end{figure}

Figure~\ref{fig:workflow_dag} shows the resulting top-level workflow DAG for a single dataset branch. Data is downloaded and partitioned into client shards, after which $T$ federated learning rounds execute as sub-workflows, each implementing the fan-out/fan-in pattern shown in Fig.~\ref{fig:flround}. A centralized baseline and result visualization complete the pipeline. The full workflow replicates this branch for each dataset, enabling parallel execution.

\subsection{Workflow Generation Comparison}
To evaluate the effectiveness of AI-assisted workflow generation, we compare four approaches: Claude Code with the \texttt{pegasus-ai} plugin, OpenAI Codex, Kimi~K2.6 running under Opencode~\cite{opencode}, and a manually implemented workflow by an expert in federated learning and Pegasus WMS (Table~\ref{tab:comparison}). All three LLM systems had access to identical documentation. Claude integrates it as structured skills invoked during generation, whereas Codex and Kimi receive it as static context in the working directory.

The comparison metrics are defined as follows. \emph{Design guidance} counts the design decisions explicitly surfaced for user approval before code generation, and \emph{orchestration options} indicates whether alternative round-orchestration strategies were presented to the user. \emph{Staging resolved} indicates whether the tool produced a working data-staging configuration for nested sub-workflows. \emph{Reproducibility} is rated \emph{high} when the workflow can be regenerated from persisted artifacts (the specification) and \emph{low} when regeneration depends on replaying a prompt history.
\begin{table}[t]
\centering
\caption{Workflow generation comparison.}
\label{tab:comparison}
\scriptsize
\setlength{\tabcolsep}{2.5pt}
\begin{tabular*}{\columnwidth}{@{\extracolsep{\fill}}p{3.0cm}cccc@{}}
\toprule
\textbf{Metric} & \begin{tabular}[c]{@{}c@{}}\textbf{Claude Code}\\\textbf{+ pegasus-ai}\end{tabular} & \textbf{Codex} & \begin{tabular}[c]{@{}c@{}}\textbf{Kimi}\\\textbf{(Opencode)}\end{tabular} & \textbf{Manual} \\
\midrule
Files generated & 22 & 33 & 15 & n/a \\
Specification & Yes & Yes & Yes & No \\
Design guidance & 5 & 0 & 0 & n/a \\
Orchestration opts. & Yes & No & No & Yes \\
\midrule
Orchestration & Sub-wf & Sub-wf$^\dagger$ & Sub-wf$^\ddagger$ & Ensemble$^\ast$ \\
Sub-workflow support & Initial & Later$^\dagger$ & Later$^\ddagger$ & Yes \\
CondorIO staging & Initial & Later$^\dagger$ & Later$^\ddagger$ & Yes \\
Parallel datasets & Initial & Later$^\dagger$ & Initial & Yes \\
GPU containers & Initial & Later$^\dagger$ & Initial$^\S$ & Yes \\
PyTorch models & Initial & Later$^\dagger$ & Initial & Yes \\
FedProx support & Initial & Later$^\dagger$ & Initial & Yes \\
Staging resolved & Yes & Yes & No$^\ddagger$ & Yes \\
\midrule
\textbf{Reproducibility} & High & Low & Low & Manual \\
\bottomrule
\end{tabular*}

\vspace{1pt}
\begin{tabular*}{\columnwidth}{@{}l@{}}
\footnotesize Sub-wf: sub-workflow. $^\dagger$Follow-up prompts required. $^\ddagger$Staging unresolved. \\
\footnotesize $^\S$Fixed in later sessions. $^\ast$Preferred for large-scale workflows.
\end{tabular*}
\end{table}

All three produced functional Pegasus workflows but differ in design completeness, interaction effort, and reproducibility. Claude Code produced a modular, feature-complete workflow~\cite{repo_claude_fl} in a single session, guided by structured skills that surface design alternatives and enforce consistent patterns. Codex produced a simplified initial workflow~\cite{repo_codex_fl} and needed repeated follow-ups to add sub-workflows, data staging, parallel dataset execution, and realistic models. Kimi, given the same documentation, required the most refinement~\cite{repo_kimi_fl} and never resolved the sub-workflow staging conflict, oscillating between flat DAG and sub-workflow approaches across several debugging sessions.

Claude also exposes orchestration alternatives before any code exists, which neither of the others does, and its specification supports regeneration from a persisted artifact where they depend on replaying an interaction history (Table~\ref{tab:comparison}).

We tested that regeneration claim rather than asserting it. The persisted specification alone was given to a fresh session with the plugin, in a working directory containing no file from the original implementation, and the criteria for equivalence were fixed before comparing. The regenerated workflow is structurally identical for the same configuration, with 15 top-level jobs, 4 round sub-workflows, and 5 jobs per round in the same fan-out and fan-in arrangement, differing only in generated file names, which the specification lists as free. It satisfies the eleven stated constraints, cites seven of them by identifier in its own source, plans without error, and reintroduces none of the nine defects recorded during the original development. Seven passages cite the originating fix by number, so the sections recording status and applied fixes are read rather than merely persisted.

\subsection{Cost of AI-Assisted Development}
\label{sec:cost}

Claude Code sessions used Claude Opus~4 (\$15/M input, \$75/M output tokens), Codex sessions used OpenAI's GPT-5.4 model, and Kimi sessions used Kimi~K2.6 via Opencode.

\begin{table}[t]
\centering
\caption{Development effort comparison. Manual figures are estimates.}
\label{tab:ai_cost}
\scriptsize
\setlength{\tabcolsep}{3pt}
\begin{tabular*}{\columnwidth}{@{\extracolsep{\fill}}lcccc@{}}
\toprule
\textbf{Metric} & \begin{tabular}[c]{@{}c@{}}\textbf{Claude Code}\\\textbf{+ pegasus-ai}\end{tabular} & \textbf{Codex} & \begin{tabular}[c]{@{}c@{}}\textbf{Kimi}\\\textbf{(Opencode)}\end{tabular} & \textbf{Manual} \\
\midrule
Sessions                 & 2            & 9            & 4            & n/a \\
Prompts, construction    & 7            & 7            & 12           & n/a \\
Prompts, refine \& debug & 45           & 42           & 51           & n/a \\
Output tokens            & $\sim$109K   & $\sim$237K   & $\sim$178K   & n/a \\
Estimated cost (USD)     & \$10 to 15   & \$15 to 20   & n/r$^\ddagger$ & n/a \\
Development time         & 2 days       & 3 weeks      & 7 days       & 3 to 4 months \\
Production-ready         & Yes          & Yes          & Partial$^\dagger$ & Yes \\
\bottomrule
\multicolumn{5}{l}{\footnotesize $^\dagger$Generates a valid DAG but not yet executed on distributed infrastructure.} \\
\multicolumn{5}{l}{\footnotesize $^\ddagger$Not recorded. Kimi ran through a gateway that reports no per-call cost.}
\end{tabular*}
\end{table}

Construction is the cheaper half for every agent, and the totals differ in where the remaining effort went rather than in its size. Codex spent its sessions incrementally adding features that Claude produced in its first, and at a comparable prompt count consumed more than twice Claude's output tokens, because it repeatedly regenerated its monolithic workflow generator. Kimi's effort was concentrated rather than spread, with a single later session consuming 51 of its 63 prompts on Pegasus-specific issues such as sub-workflow file registration and staging policy conflicts, which it never resolved. Refinement and debugging dominated Claude's effort as well, at 45 of 52 prompts, though from a far lower base and against a workflow that already ran.

Every LLM-assisted approach costs negligibly little against the GPU resources the experiments required, nine GPUs for multiple days. In Table~\ref{tab:ai_cost}, \emph{production-ready} means the generated workflow executed end to end on the pool of Section~\ref{sec:experiments}, and the manual figure is the prior-work estimate cited in the introduction.
\subsection{AI Debugging Skill}
\label{sec:debugging}

Across the runs the \texttt{/pegasus-debug} skill resolved failures spanning Pegasus configuration, HTCondor scheduling, Docker containerization, and filesystem capacity. The loop of Section~\ref{sec:architecture} ran autonomously throughout, without per-fix human approval, and human involvement was limited to post-hoc review of the applied changes.

What it resolved ranged across all four of those layers. \emph{Missing cleanup} let intermediate files accumulate across rounds until in-place cleanup was enabled. \emph{Experiment collisions} occurred when concurrent experiments overwrote shared output paths, fixed by per-experiment namespacing. A \emph{model configuration error}, a mismatch between model output dimensions and class-weight computation, failed training jobs until the loss configuration was corrected. \emph{Missing dependencies}, undeclared script inputs, were added to the workflow definition.

\paragraph{What the errors have in common} Rather than count classes, we classified every change the specification records as applied during development, which separates nine genuine defects from five features and design changes. We organize them not by the subsystem they belong to but by whether the runtime produced any evidence to diagnose (Table~\ref{tab:defects}). Six of the nine produced such evidence: a job failed, and the exit codes, logs, and held-job records that a debugging loop consumes were all present. The other three left no trace at all. The workflow ran to completion, every job exited zero and wrote its declared outputs, and the results were nonetheless wrong.

This division comes from the execution logs rather than from our judgment, and it decides what any failure-driven method can reach. A debugging loop covers the first six by construction and cannot reach the other three. What the three silent defects share is a common shape rather than a common subsystem. In each case the workflow did exactly what it had been told, and what it had been told was wrong: a shard was read from a directory the partitioning job never wrote, an algorithm parameter never reached the trainer, and an output path was shared by concurrent configurations. The picture reverses when we ask what was visible in the generated source. All three silent defects were fully visible there before submission, whereas two of the six that failed appeared only as anti-patterns whose trigger depends on worker disk capacity or on the class distribution of a shard. Static checking is therefore most reliable exactly where the runtime is blind. This result bounds what a pre-submission check could reach. It does not measure what a particular checker finds, because the false positive rate of such a pass is not evaluated here. The per-defect classification is released with the workflow~\cite{repo_claude_fl}.

\begin{table}[t]
\centering
\caption{The nine defects by runtime evidence. Statically visible counts evidence in the generated source before submission.}
\label{tab:defects}
\scriptsize
\setlength{\tabcolsep}{4pt}
\begin{tabular*}{\columnwidth}{@{\extracolsep{\fill}}lccc@{}}
\toprule
\textbf{Runtime evidence} & \textbf{Defects} & \textbf{Debug loop} & \textbf{Statically visible} \\
\midrule
A job failed   & 6 & reaches & 4 full, 2 partial \\
Nothing failed & 3 & cannot reach & 3 full, 0 partial \\
\midrule
Total          & 9 &  & 7 full, 2 partial \\
\bottomrule
\end{tabular*}
\end{table}

\paragraph{Worked example: disk exhaustion} One of the six failures shows the debugging loop end to end, and it also shows why a workflow failure is not the same problem as a program failure. Midway through a fifty round run, training jobs began failing on workers whose scratch space had filled. No single layer of the system explains this. The job logs say only that training died. The cause lies in how the planner staged each round, not in anything a job did. The symptom appears on a worker machine that neither the logs nor the planner names. The agent read across all three layers and found the mechanism: every round's sub-workflow was staging its own copy of the multi-gigabyte training container, so disk use on each worker grew with the round count and crossed the limit partway through the run. The correction is a single property on the container definition, which tells workers to pull the image once and reuse their local cache. The agent applied it to the workflow generator, re-planned, and re-submitted the affected sub-workflows, which then completed. No human approved any step.

The episode also shows what the specification adds to the repair. The property is now recorded there as a constraint, so it can be checked before submission, and a regenerated workflow carries it by construction rather than by anyone's memory of the incident.
\paragraph{A defect that produced no failure} One of the nine defects was found only after the experiments had been reported. The client training script read its data shard from a directory whose name did not match the one the partitioning job wrote. The dataset class therefore found no images and substituted random tensors, so every training job trained on noise. Counted from the Kickstart records, this affected all 1,700 training jobs in the three reported configurations, and 3,700 across every configuration run during development. Evaluation and the centralized baselines read real data, because their paths happened to match. The reported federated accuracies therefore came from noise-trained models evaluated on a real test set, which is why they sat near the majority-class rate while the centralized baselines did not.

The metrics these jobs produced were low but unremarkable for federated learning on heterogeneous medical images, so nothing looked wrong and the failure-driven loop had nothing to act on. The defect was found instead by a conformance pass, in which ten independent reviewers, given only the specification and the code, judged the implementation against the constraints of Section~\ref{sec:spec_driven}. All ten flagged the constraint forbidding synthetic-data substitution, and four of them independently identified the directory mismatch as the reason the forbidden path was always taken. When the pass was repeated on the fixed code, every verdict it returned could be confirmed in the source, and it surfaced three further genuine violations of stated constraints. On a single codebase this is an observation rather than a measured false-positive rate.

Two qualifications apply. First, the constraint in question names the forbidden construct explicitly, so the reviewers were in effect told what to look for. The result is evidence that checking a written constraint is what mattered, not evidence of unaided discovery. The constraint had been written weeks earlier and changed nothing until it was checked. Second, the reviewers were ten sessions of a single model, so their agreement is less independent than the count suggests. Only six of the nine defects recorded during development fall under a stated constraint at all, which bounds what this form of checking can reach.

\paragraph{LLM vs.\ user vs.\ expert}
A non-expert user typically builds a sequential pipeline with predefined rounds and explicit job definitions and no advanced orchestration. An expert judges the generated workflow structurally correct, though FL-specific refinements such as early stopping or Ensemble Manager orchestration~\cite{safri2024workflow} are absent.

\subsection{Federated Learning Performance}

Model quality is not a contribution of this work, and these results only confirm that the generated workflows execute correctly. They are re-runs after the one line fix for the silent data-path defect of Section~\ref{sec:debugging}, which invalidated an earlier version of them.

On the corrected runs, federated learning approaches centralized training on both datasets
(Table~\ref{tab:fl_accuracy}). Convergence is rapid, with TCIA passing 90\% by round four and plateauing near
93\% by round twenty. Reducing the client count (E3) narrows the gap to under a point on TCIA and closes it on
NIH. Reducing rounds while raising local epochs (E4) matches the fifty round baseline in a fifth of the rounds
and 4.6 hours instead of 24.3, at one fifth of the model exchanges. These numbers match the expected behaviour of FedAvg on naturally partitioned data rather than a heterogeneity penalty. %E2 had not finished re-running when the testbed entered maintenance, so no FedProx result is reported.

\begin{table}[h]
\centering
\caption{Final-round accuracy vs.\ the centralized baseline, percent.}
\label{tab:fl_accuracy}
\scriptsize
\begin{tabular*}{\columnwidth}{@{\extracolsep{\fill}}lcccc@{}}
\toprule
 & \multicolumn{2}{c}{\textbf{TCIA}} & \multicolumn{2}{c}{\textbf{NIH ChestX-ray}} \\
\cmidrule(lr){2-3} \cmidrule(lr){4-5}
\textbf{Exp} & Federated & Centralized & Federated & Centralized \\
\midrule
E1 & 93.6 & 95.8 & 66.1 & 67.4 \\
E3 & 95.3 & 95.9 & 66.8 & 66.7 \\
E4 & 94.2 & 95.1 & 67.6 & 66.3 \\
\bottomrule
\end{tabular*}
\end{table}

\subsection{Workflow Execution Statistics}
The fifty round experiments consist of 101 sub-workflows, 50 rounds across two datasets plus the top level, and generate over 2,000 jobs each (Table~\ref{tab:execution_stats}).
\begin{table}[h]
\centering
\caption{Structure and execution of the corrected runs. Occ. is the fraction of a median round's wall time with a job executing.}
\label{tab:execution_stats}
\scriptsize
\setlength{\tabcolsep}{4pt}
\begin{tabular*}{\columnwidth}{@{\extracolsep{\fill}}lccc ccc ccc@{}}
\toprule
 & \multicolumn{3}{c}{\textbf{Configuration}} 
 & \multicolumn{3}{c}{\textbf{FL Structure}} 
 & \multicolumn{3}{c}{\textbf{Execution}} \\
\cmidrule(lr){2-4} \cmidrule(lr){5-7} \cmidrule(lr){8-10}
\textbf{Exp} & K & T & E & Train & Agg & Eval & Jobs & Wall (h) & Occ. \\
\midrule
E1 & 10 & 50 & 2 & 1000 & 100 & 4 & 2679 & 24.3 & 12.8\% \\
E3 & 5  & 50 & 2 & 500  & 100 & 4 & 2179 & 30.1 & 8.0\% \\
E4 & 10 & 10 & 5 & 200  & 20  & 4 & 559  & 4.6  & 27.9\% \\
\bottomrule
\end{tabular*}
\end{table}
Reducing rounds (E4) cuts workflow complexity by 80\% and execution time to 4.6 hours. Reducing clients (E3) lowers training jobs but not the number of sub-workflows, and it lengthens the run to 30.1 hours because each of the five remaining clients then carries twice the data, so halving the client count redistributes work rather than removing it.

The job-level provenance records show where the remaining time goes. Within a median E1 round of 1,601 seconds, at least one job is executing for 204 seconds and DAGMan startup and teardown account for 13. The remainder is dominated by data movement rather than by orchestration. Under CondorIO every job receives its inputs afresh, so a median client training job executes for 17.7 seconds while a further 134 seconds of its slot is spent outside process execution, during which HTCondor moves 75 MB in and 45 MB out. Jobs that receive no large inputs show no comparable gap, which is what attributes it to transfer rather than to generic slot overhead. A median training job also waits 361 seconds between submission and execution, because a round requests ten GPUs per dataset branch from a pool of nine. Each round moves about 2.6 GB, and a fifty round experiment roughly 260 GB, largely because the same client shards and global model are re-staged on every round.

Round occupancy therefore tracks the amount of computation a round contains rather than its size. E4, with five local epochs per round, reaches 27.9\% against 12.8\% for E1 and 8.0\% for E3. Expressing each round as a sub-workflow costs little in itself, and the case for doing so strengthens as per-round computation grows relative to the fixed cost of staging its inputs. All three corrected configurations completed without job-level retries and without DAG-level rescues.

%============================================================
\section{Related Work}
\label{sec:background}

Several systems generate workflows from natural-language input. Zeng et al.~\cite{zeng2024flowmind} use structured prompting and API grounding, and bioinformatics studies show similar applications in Galaxy and Nextflow~\cite{prompt2025pipeline}. Zhang et al.~\cite{zhang2024aflow} model generation as a search over tool graphs, which is not tailored to DAG-based scientific workflows and lacks WMS integration, and Balis et al.~\cite{balis2026workflow} combine intent extraction with deterministic DAG construction but target HyperFlow and do not address distributed orchestration. None of these combines workflow synthesis with execution in a system such as Pegasus.

On the debugging side, code-generation models such as Codex~\cite{chen2021evaluating}, multi-agent frameworks such as AutoGen~\cite{AutoGen2024}, and self-debugging techniques that repair code from execution feedback~\cite{chen2023selfdebug} all operate on standalone programs, whereas diagnosing a workflow failure requires interpreting evidence spread across the WMS, the scheduler, and the container runtime. Within workflow systems, PANORAMA~\cite{deelman2017panorama} combines performance models with online monitoring to diagnose runtime anomalies in Pegasus workflows, but leaves the corrective action to the operator. The debugging skill addresses both gaps: it reasons across system layers and applies the fix.

Federated learning also has dedicated orchestration frameworks such as Flower~\cite{beutel2020flower} and PySyft~\cite{ziller2021pysyft}, which manage training rounds but lack the provenance, data staging, and fault tolerance of a scientific WMS. Safri et al.~\cite{safri2024workflow} expressed FL as manually constructed Pegasus workflows. Our work generates such workflows automatically from natural-language intent.

%============================================================

\section{Conclusion}
\label{sec:conclusion}
We studied the composition and management of a federated learning workflow for medical imaging under AI assistance, following it from natural-language intent through a reviewable specification, code generation, autonomous debugging, and execution on distributed infrastructure. Separating design from implementation made the rest possible: the specification was reviewable while the code was still hypothetical, and it later proved detailed enough both to audit that code and to rebuild it. %Separating design from implementation is what makes the rest work. The specification was reviewable while the code was still hypothetical, and it later proved detailed enough both to audit that code and to rebuild it.

The generated workflow reached 101 sub-workflows and 2,679 jobs on nine GPUs across four sites, a design that an expert puts at 3 to 4 months of manual development and that was produced here in two days. Among the LLM agents evaluated, the one grounded in structured skills surfaced orchestration alternatives before generating any code and resolved sub-workflow staging on its first attempt, while agents given the same documentation as unstructured context reached a working design only through repeated follow-up prompts, or not at all.

%The defects encountered tell the sharpest story. Six of the nine announced themselves as job failures, which is the evidence a debugging loop runs on. The other three produced no failure at all, among them one that trained every one of 1,700 training jobs on random tensors while each job exited successfully and wrote well formed metrics. All three were fully visible in the generated source, and checking the specification's constraints against that source is what found them. 
Six of the nine defects surfaced as job failures, the evidence a debugging loop runs on. The other three produced no failure at all, including one that trained all 1,700 training jobs on random tensors while every job exited successfully with well formed metrics. All three were visible in the generated source, and checking the specification's constraints against it found them. Static checking is most reliable exactly where the runtime is blind, so failure-driven repair and specification-level checking reach disjoint classes of error and a validation loop for AI-generated workflows needs both rather than either alone.

Our evaluation covers one application and one workflow management system. Future work will apply the methodology to other scientific workflows, workflow systems, and AI coding agents to assess its generality.
%Although our evaluation focuses on a single application and workflow management system, future work will evaluate the proposed methodology across a broader range of scientific workflows, workflow management systems, and AI coding agents to further assess its scalability and generality.

% ============================================================
\section*{Acknowledgments}
This work is supported by NSF grant \#2513101 and the FABRIC Testbed (NSF \#2330891). The authors used Claude to help with the pictures. All intellectual contributions and final editorial decisions remain solely with the human authors.
%This work is supported by US National Science Foundation grant \#2513101, and we acknowledge the FABRIC Testbed (NSF \#2330891). %The authors used Claude to help edit the manuscript. All intellectual contributions and final editorial decisions remain solely with the human authors.
\balance
\bibliographystyle{IEEEtran}
\bibliography{references}

\end{document}